\documentclass[superscriptaddress,floatfix,twocolumn,amsmath,amssymb]{revtex4-1}

\usepackage{tabu}
\usepackage{gensymb}
\usepackage{graphicx}
\usepackage{microtype}
\usepackage{dcolumn}
\usepackage{bm}
\usepackage{amssymb}
\usepackage{natbib}
\usepackage{mathtools}
\usepackage{color}
\usepackage{datetime}
\usepackage{footnote}
\usepackage{bbold}
\usepackage[normalem]{ulem}
 \usepackage{ragged2e}
 \usepackage{xfrac}
 \usepackage{sidecap}
 \usepackage{setspace}
 \usepackage{multirow}
\usepackage{xcolor}
\usepackage{hyperref}
\usepackage{cleveref}
\usepackage{siunitx}
\usepackage[utf8]{inputenc}

\begin{document}

\title{High coupling efficiency grating couplers on lithium niobate on insulator}

\author{Inna Krasnokutska}
\affiliation{Quantum Photonics Laboratory and Centre for Quantum Computation and Communication Technology, School of Engineering, RMIT University, Melbourne, Victoria 3000, Australia}

\author{Robert J. Chapman}
\affiliation{Quantum Photonics Laboratory and Centre for Quantum Computation and Communication Technology, School of Engineering, RMIT University, Melbourne, Victoria 3000, Australia}

\author{Jean-Luc J. Tambasco}
\affiliation{Quantum Photonics Laboratory and Centre for Quantum Computation and Communication Technology, School of Engineering, RMIT University, Melbourne, Victoria 3000, Australia}

\author{Alberto Peruzzo}
\thanks{alberto.peruzzo@rmit.edu.au}
\affiliation{Quantum Photonics Laboratory and Centre for Quantum Computation and Communication Technology, School of Engineering, RMIT University, Melbourne, Victoria 3000, Australia}

\begin{abstract}
We demonstrate monolithically defined grating couplers in $Z$-cut lithium niobate on insulator for efficient vertical coupling between an optical fiber and a single mode waveguide. The grating couplers exhibit $\sim 44.6\%$/coupler and $\sim 19.4\%$/coupler coupling efficiency for TE and TM polarized light respectively. Taperless grating couplers are investigated to realize a more compact design.
\end{abstract}

\maketitle

\section{Introduction}

Lithium niobate on insulator (LNOI) is emerging as a promising platform for photonic technology.
Recent works highlight the potential of the high index contrast lithium niobate
platform through low-loss waveguides \cite{Zhang:17}, optical modulators \cite{WangNat:18, Mercante:18},
tunable ring resonators \cite{Wang:18, krasnokutska2018large,Siew:cx} and nonlinear effects \cite{Wangmg:18}.  This
platform has potential to make a mark in fields ranging from telecommunications \cite{WangNat:18, Mercante:18}
to quantum computing and communication \cite{Aghaeimeibodi:18} as well as sensing \cite{Jiang:2016fl}; however,
many challenges are yet to be overcome.

A major obstacle hampering the long-term success of photonic platforms is the
large fiber-to-chip coupling loss.  Ultimately, compact, low-loss, wavelength
and polarization insensitive in--out couplers are needed to improve device
efficiency.  Much progress has been made with silicon photonics, including specialized inverse taper \cite{Fu:14}, metastructured
butt-couplers \cite{Shen:14} and low index contrast overlays using materials such as SiON and
SU-8 \cite{Park:13,Pu:2010,WAHLBRINK20091117}.  Recent works have investigated
efficient butt-coupling into LNOI \cite{he2019low,krasnokutska2019nanostructuring}; however, this technique is particularly sensitive to fiber misalignment and requires the waveguides to extend to the facet of the chip, often consuming valuable chip realestate.
Grating couplers allow coupling anywhere inside a chip with minimal fiber alignment
tolerance \cite{Baets}; this is critical for the efficient mass characterization of devices
across a wafer and useful for specific fixed-wavelength and fixed-polarization devices.

Grating couplers can be challenging structures to fabricate as they consist of
narrow and closely spaced ridges.  Significant works in Si and SiN to optimize grating couplers via apodizing
the gratings \cite{Ding:14}, using underlying reflective stacks \cite{Sodagar:14}
and curving the gratings \cite{imecgc} have been demonstrated.
Minimal grating coupler work in LNOI has been demonstrated to date, possibly
due to the challenges in nanostructuring lithium niobate \cite{Chen:17, Aghaeimeibodi:18, Jian:18}.
Focused ion beam (FIB) was used to demonstrate proof-of-principle grating couplers in LNOI \cite{Chen:17},
and Si on LNOI gratings \cite{Jian:18} were presented, and recently, gratings
from etched LNOI were shown \cite{Aghaeimeibodi:18}. Unfortunately, grating couplers in LNOI reported to date are affected by low coupling efficiency.

In this work, we demonstrate high efficiency monolithically defined C-/L-band (1520--1630\,nm wavelength)
grating couplers in $Z$-cut LNOI with efficiency of $\sim 44.6\%$/coupler for TE polarization,
and $\sim 19.4\%$/coupler for TM polarization.  Grating coupler design parameters
are investigated through simulation and experiment, and the performance of compact
(taperless) grating couplers is evaluated.  We anticipate these results to assist in the production of efficient LNOI devices.

\section{Design and fabrication}

\begin{figure*}[t]
\centering
\includegraphics[width=1.0\linewidth]{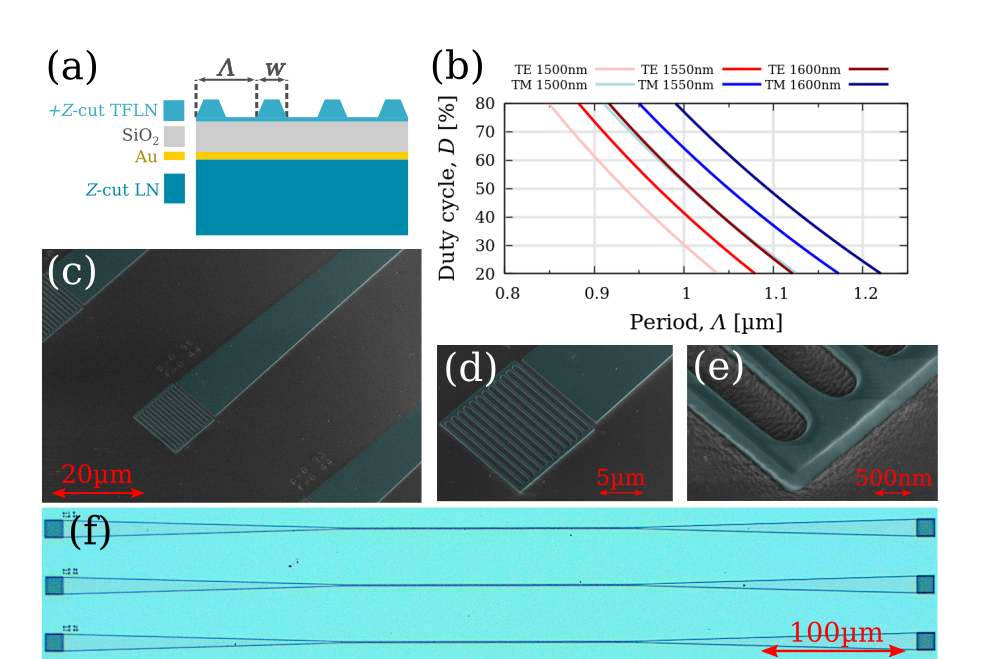}
\caption{(a) A schematic representation of the fabricated sample cross-section, where $\Lambda$ is the grating coupler period, $w$ is the tooth width and $D=w/\Lambda$ is the grating duty cycle. (b) Semi-analytic calculation of the grating period and duty cycle for TE and TM polarization to couple 1500\,nm, 1550\,nm and 1600\,nm wavelengths. Scanning electron microscope images of fabricated grating couplers on $Z$-cut LNOI taken at a $40^{\circ}$ tilt with light-blue false coloring to highlight the fabricated structures: (c) grating couplers with mode-matching tapers; (d) close-up of an etched grating with period $\Lambda=0.96$\,$\mu$m and duty cycle $D=40\%$; (e) close-up of the lower corner of a grating coupler. (f) An optical micrograph showing three entire grating coupler structures.}
\label{fig1}
\end{figure*}

We designed and fabricated C-/L-band grating couplers on $Z$-cut LNOI. The schematic representation of the grating cross section is shown in Fig.\ref{fig1}(a) 500\,nm thick $Z$-cut LN thin film (TFLN) atop a 2\,$\mu$m thick buried SiO$_2$ layer and 100\,nm gold (Au) layer supported by a single-crystal $Z$-cut LN substrate. The gold layer was already present in the wafer for use as a ground electrode for electrically tunable ring resonators \cite{krasnokutska2018large}. The sample was patterned and etched 400\,nm deep using the process reported in \cite{Krasnokutska:18}. The first order ($O=1$) grating period ($\Lambda$) and duty cycle values ($D$) presented in Fig.\ref{fig1}(b) are estimated semi-analytically for TE and TM polarized light with wavelengths 1500\,nm, 1550\,nm and 1600\,nm via
\begin{align}
\Lambda = \frac{2 \pi O}{\beta - k_0 n_\mathrm{inc} \sin(\theta_\mathrm{inc})}
\end{align}
where $k_0 = 2\pi / \lambda_0$ is the free space wavenumber and $\lambda_0$ is the corresponding operating wavelength, $\beta = n_\mathrm{eff} k_0$ is the propagation constant, and $n_\mathrm{inc} = 1$ is the refractive index of the cladding medium, which in this case is air \cite{Baets}.
The effective index, $n_\mathrm{eff}$, was found using a fully vectorial mode solver taking into account the wavelength and temperature dependent anisotropic refractive index of lithium niobate.  All designs are based on an incident fiber angle of $\theta_\mathrm{inc}=8^\circ$.

Based on the simulation results in Fig.\ref{fig1}(b), grating coupler structures with periods ranging from 0.92 to 1\,$\mu$m (TE operation) and 1 to 1.08\,$\mu$m (TM operation) were fabricated.  These structures consist of two $12\times10$\,$\mu$m grating couplers to couple light in and out of the chip, two mode-matching tapers with widths linearly varying from 12 to 1\,$\mu$m over a 200\,$\mu$m length, and a 200\,$\mu$m long and 1\,$\mu$m wide waveguide. An optical micrograph showing three complete test structures is presented in Fig.\ref{fig1}(f), while Figs.\ref{fig1}(c)-(e) show false colored scanning electron microscope (SEM) images of a grating coupler at various magnification levels.

\section{Results and Discussion}
\begin{figure*}[t]
\centering
\includegraphics[width=1.0\linewidth]{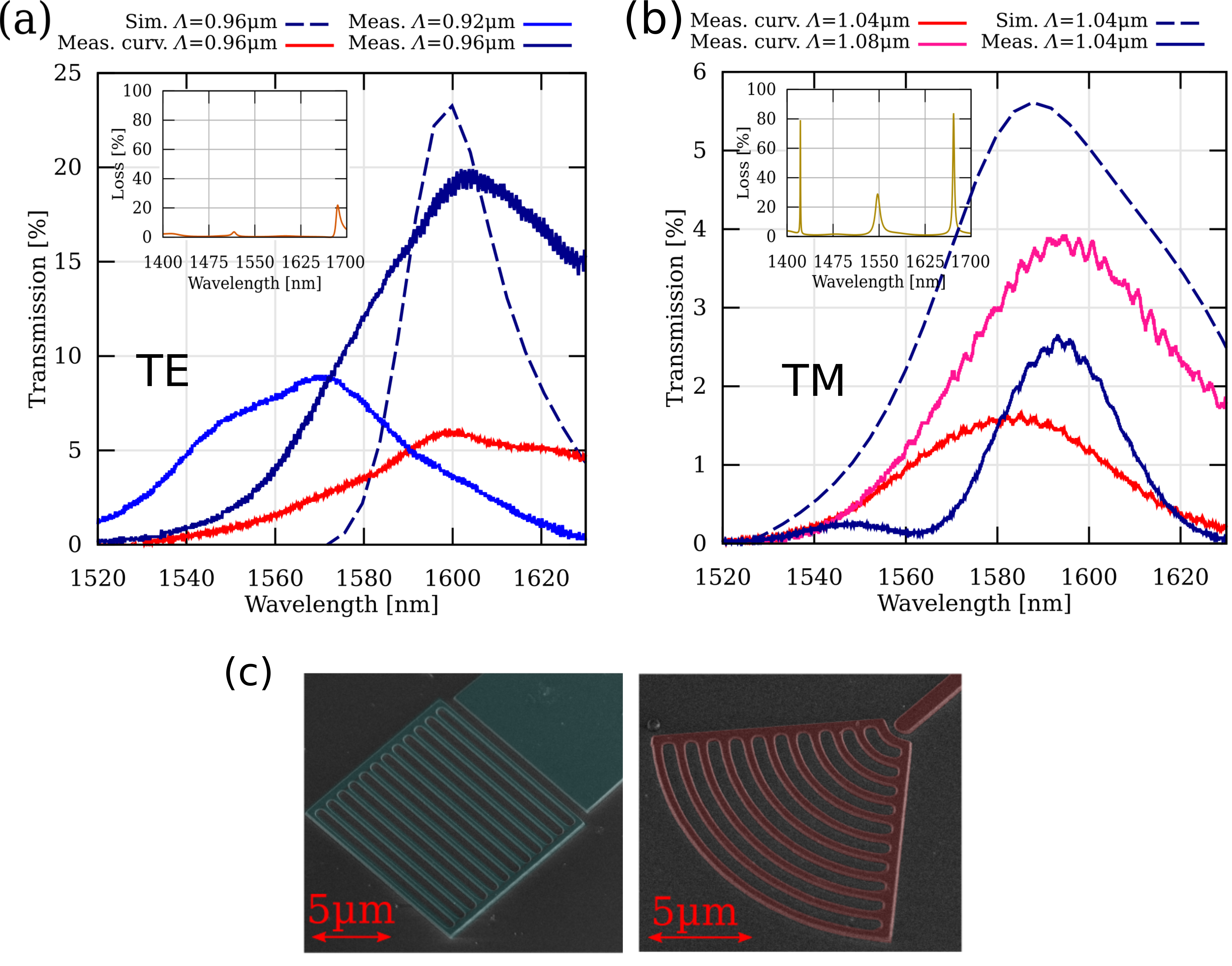}
\caption{Simulated and measured transmission spectrum of the grating couplers for the TE (a), and TM (b) polarization. Simulations of the induced losses due to gold layer cavity are reported in the insets. Data for the straight (curved) gratings are plot in blue (red) color. (c) False-color SEM pictures of straight and curved gratings.}
\label{fig2}
\end{figure*}

Two SMF-28 fibers at an $8^{\circ}$ angle have been used to couple light in and out of the grating couplers. The wavelength response of the structures has been recorded by continuously sweeping the laser wavelength from 1520 to 1630\,nm and measuring the overall transmission through the device. The transmission has then been normalized to the wavelength spectrum of the laser. We assume that the propagation loss in the straight waveguides and in the inverse tapers is negligible based on our previous experiments \cite{Krasnokutska:18} and that in and out grating couplers have similar performance. The transmission ($T$) can therefore be used to calculate the efficiency of each coupler as $\eta=\sqrt{T}$ \cite{Baets}.

Fig.\ref{fig2}(a) shows the measured and simulated transmission spectrum of the grating couplers with $\Lambda$=0.92 and 0.96\,$\mu$m and $D=40\%$ for TE polarized light. The dashed blue curve is the simulated coupling efficiency computed by Lumerical FDTD. We achieve 20\,$\%$ fiber-to-fiber transmission corresponding to a grating coupler efficiency of $\sim 44.6\%$/coupler, which is 4\,$\%$ less than predicted by simulations. Using a curved grating coupler, which does not require a long inverse taper and is shown in a red colored SEM image in Fig.\ref{fig2}(c), we observe a coupling efficiency of $\sim 24\%$/coupler. In Fig. \ref{fig2}(b) we report TM transmission measurements using grating couplers with periods $\Lambda =1.04$ and 1.08\,$\mu$m and duty-cycle $D=55\%$; the corresponding dashed-blue curve highlights the simulation result. We implement compact curved grating couplers with the same period and duty-cycle, and observe lower transmission mainly due to the mode mismatch between the grating coupler and the single mode waveguide. However, for the curved grating with $\Lambda =$1.08\,$\mu$m we achieve an overall transmission of 3.8\,$\%$ and a coupling efficiency of 19.4\,$\%$/coupler. 

To better understand the transmission spectrum, we further investigated the influence of the gold bottom layer on the grating coupler performance using a Fourier modal method (FMM) film analysis simulator, $S^4$. We found that due to the presence and position of the gold layer, the stack in Fig. \ref{fig1}(a) forms a resonant cavity around 1550\,nm wavelength for TM causing gold absorption losses; the insets to Fig. \ref{fig2}(a) and (b) show the simulated absorption spectrum for TE and TM polarized light. We observe a transmission dip around 1550\,nm wavelength for the experimental TM measurements of the straight grating couplers with a period of $\Lambda =$1.04\,$\mu$m that can be associated to cavity losses caused by the material stack. We note that material stack absorption should not impact the performance of the grating couplers operating under TE polarized light, as can be seen from the inset to Fig. \ref{fig2}(a). The material stack cavity losses can be mitigated by shifting the cavity resonance away from the grating coupler operating wavelength by potentially adjusting the film thickness of the LN, the thickness of the SiO$_2$ bottom cladding, or the grating coupler etch depth.

\section{Conclusion}
We have presented record high-efficiency grating couplers fabricated using reactive ion etching (RIE) in $Z$-cut LNOI.
Semi-analytical modelling, FDTD modelling and FMM film analysis were used to design and confirm the performance of the fabricated grating couplers.
A maximum measured transmission of $\sim 44.6\%$/coupler and $\sim 19.4\%$/coupler for TE and TM grating couplers respectively was achieved. We anticipate these results to facilitate the development of complex LNOI circuits, in particular for mass-characterization of photonic components, simplified fiber-to-chip coupling and compact device design. 

\section*{Funding}
Australian Research Council Centre for Quantum Computation and Communication Technology CE170100012; Australian Research Council Discovery Early Career Researcher Award, Project No. DE140101700; RMIT University Vice-Chancellors Senior Research Fellowship.

\section*{Acknowledgments}
 This work was performed in part at the Melbourne Centre for Nanofabrication in the Victorian Node of the Australian National Fabrication Facility (ANFF) and the Nanolab at Swinburne University of Technology. The authors acknowledge the facilities, and the scientific and technical assistance, of the Australian Microscopy \& Microanalysis Research Facility at RMIT University.

\end{document}